\documentclass[twoside,12pt]{article}
\usepackage{a4,amsmath,amssymb,amsfonts}

\title{\bf On the unitarity problem in space/time noncommutative theories}

\author{D. Bahns\thanks{Supported by the Graduiertenkolleg ``Zuk\"unftige Entwicklungen in der Teilchenphysik''.} , 
S. Doplicher\thanks{Research supported by MIUR and GNAMPA - INdAM.} , 
K. Fredenhagen$^\ddag$, G. Piacitelli$^\S$
\\\\
$
\begin{array}{l}
\!\!\phantom{ }^* \phantom{ }^\ddag
\mbox{{\footnotesize II. Institut f\"ur Theoretische Physik, Universit\"at
Hamburg, Luruper Chaussee 149, }}
\\ \phantom{^*}\mbox{{\footnotesize D - 22761 Hamburg, Germany --
 {\tt bahns@mail.desy.de, fredenha@x4u2.desy.de}}}
\\^\dagger\mbox{{\footnotesize Dipartimento di Matematica
, Universit\`a di Roma ``La Sapienza'',
P.le Aldo Moro 2, }}
\\\phantom{^\dag}\mbox{{\footnotesize 00185 Roma, Italy --{
\tt dopliche@mat.uniroma1.it}}}
\\^\S\mbox{{\footnotesize Dipartimento di Matematica Pura ed Applicata,
Universit\`a di Padova,}} 
\\\phantom{^\S}\mbox{{\footnotesize  
Via G. Belzoni 7, 35131 Padova, Italy -- {\tt piacitel@math.unipd.it}}}
\end{array}$}

\date{}

\begin{document}
\maketitle

\begin{abstract}
\noindent It is shown that the violation of unitarity observed in space/time
noncommutative field theories is due to an improper definition of quantum field
theory on noncommutative spacetime.
\end{abstract}

\noindent Quantum field theory on noncommutative spacetimes is an interesting
modification of the standard formalism which takes into account possible
deviations from the smoothness of spacetime at small distances.  A very
convincing reason for such a modification is that uncertainty relations for
spacetime coordinates are suggested by an analysis of the principles of quantum
theory together with those of general relativity. Such uncertainty relations
can be implemented in the framework of noncommutative spaces~\cite{dfr}. A
further motivation was established in the context of string theory in the
analysis of D-branes in background magnetic fields \cite{schomerus}.

Quantum field theory on the standard noncommutative spacetime is equivalent to
a nonlocal theory on a commutative spacetime~\cite[$\!\!\S$ \,6]{dfr}. Owing to
the nonlocality especially in time, certain structural properties of ordinary
quantum field theories are lost. In particular, the various equivalent
formulations of quantum field theory on Minkowski space cease to be equivalent
on noncommutative spaces.

Most of the current literature on the subject is based on a set of modified
Feynman rules first formulated in~\cite{filk}. The basic idea in this
approach is that the usual Feynman rules of quantum field theory on Minkowski
space may be used and that the transition to a noncommutative spacetime is
achieved by a mere replacement of all pointwise products by Moyal
star products and subsequent symmetrization. In momentum space this rule
entails the appearance of oscillating factors at the vertices, which depend on
the order of the in- and out-going momenta.

The formalism unfortunately suffers from a violation of unitarity, as was first
remarked in~\cite{gomis} (cf. also~\cite{alvarez,luk}). As proposed 
there, let us consider as an illustrative example the fish graph
in a model with $\phi^3$ self-interaction in the setting of the modified
Feynman rules, where
$$
\begin{picture}(70,23)(-3,10)
\put(00,13){\line(3,0){20}}
\put(20,13){\circle{2}}
\put(40,13){\circle{2}}
\put(40,13){\line(3,0){20}}
\put(30,13){\circle{20}}
\put(1,16){\footnotesize{$p$}}
\put(20,28){\footnotesize{$p-k$}}
\put(28,-10){\footnotesize{$k$}}
\put(55,16){\footnotesize{$p$}}
\put(7,11.5){\tiny{$>$}}
\put(45,11.5){\tiny{$>$}}
\put(26,21.25){\tiny{$>$}}
\put(26,1.75){\tiny{$>$}}
\end{picture}
\quad\propto\quad 
\int d^4k \frac{1+\cos p \,\sigma
k}{((p-k)^2-m^2+i\varepsilon)(k^2-m^2+i\varepsilon)}
$$
with $i\sigma^{\mu\nu}=[x^\mu \stackrel{\displaystyle*_\sigma}{,} x^\nu]$
defining the Moyal star product $f*_\sigma g(x)=
e^{\frac{i}{2}\stackrel{\leftarrow}{\partial}_\mu\sigma^{\mu\nu}
\stackrel{\rightarrow}{\partial}_\nu}f(x)g(x)$. 
The above expression for the fish graph is the Fourier transform of a pointwise
product and a star product of two Feynman propagators,
$$
\Delta_F(x)\Delta_F(x)+
\Delta_F(x)*\Delta_F(x)\,.
$$
Here, as well as in the sequel,
the $*$ denotes the Moyal star product defined by~$2\sigma$, whereas the
symbol $*_\sigma$ is reserved for the star product defined by~$\sigma$.

The first term in the above expresssion, also referred to as the planar
contribution, is the same as in the commutative case and thus satisfies
unitarity,
\begin{eqnarray}\label{uni}
\Delta_F^2+\overline{\phantom{\big|}\hspace{-4pt}\Delta_F^2}
&=&
\left(\theta\Delta_+ +(1-\theta)\Delta_-\right)^2
+\left(\theta \Delta_- +(1-\theta)\Delta_+\right)^2\nonumber\\
&=&\theta \Delta_+^2 +(1-\theta)\Delta_-^2
+\theta \Delta_-^2 +(1-\theta)\Delta_+^2\nonumber\\
&=&\Delta_-^2+\Delta_+^2\,,
\end{eqnarray}
where $\theta\Delta_\pm$ abbreviates the product $\theta(x_0) \Delta_\pm(x)$ of
Heaviside function and positive/negative frequency parts of the propagator. On the contrary, the analogous calculation for the second term, the ``non-planar
contribution'', yields (since the Moyal star product is Hermitean)
\begin{eqnarray}\label{nonunistar}
\Delta_F*\Delta_F +
\overline{\phantom{\big|}\hspace{-4pt}\Delta_F*
\Delta_F} 
&=&
\left(i\theta\Delta +\Delta_-\right)
*\left(i\theta\Delta +\Delta_-\right)
+\left(-i\theta \Delta +\Delta_+\right)
*\left(-i\theta \Delta +\Delta_+\right)\nonumber\\
&=&\Delta_+*\Delta_++\Delta_-*\Delta_- +
\nonumber \\&&+\,
i\,\theta\Delta*(i\theta\Delta-i\Delta)
+\,(i\theta\Delta-i\Delta)* i\,\theta\Delta
\nonumber\\
&=&\Delta_+*\Delta_++\Delta_-*\Delta_- +
\Delta_{ret}*\Delta_{av}
+\Delta_{av}* \Delta_{ret}\,,
\end{eqnarray}
with $i\Delta$ denoting the commutator function $\Delta_+-\Delta_-$, and 
$\Delta_{ret/ av}(x)=\theta(\pm\, x_0)\Delta(x)$ the retarded and the advanced
propagator, respectively. The violation of unitarity is therefore due to the
fact that in general, the (Fourier transform of the) star product of retarded
and advanced propagator does not vanish. 

These terms disappear, however, when the time is assumed to commute with
the space variables, 
i.e. when there is a timelike vector 
$n$ with $\sigma^{\mu\nu}n_{\nu}=0$. In these cases we have
$$
\Delta_{ret/av}(x)=\theta(\pm nx)\Delta(x)=\theta(\pm nx)*\Delta(x)
$$
and hence 
$$
\Delta_{ret}*\Delta_{av}=\Delta*\theta*(1-\theta)*\Delta
$$
and
$$
\theta*(1-\theta)=\theta(1-\theta)=0 \ .
$$
By continuity, this remains true when $n$ approaches a 
lightlike vector. This situation has been termed lightlike 
noncommutativity in~\cite{gomis2}. It occurs as a scaling limit 
for a generic $\sigma$.

In~\cite{dfr} a different definition of a scalar field theory on noncommutative
spacetime has been proposed, which does lead to a unitary S-matrix. It is based
on the introduction of an interaction Hamiltonian in Fock space, defined as 
$$
H_I(t)=\int\limits_{x_0=t} \hspace{-8pt} d^3x \;
:\!\phi*_\sigma\dots*_\sigma \phi (x)\!:
$$
where the integration at $x_0=t$ is given a precise meaning as a positive map.
Since $H_I(t)$ is  formally self-adjoint, the corresponding perturbative
expansion must be formally unitary. 

For the sake of  completeness  let us also mention that in~\cite{dfr} the value
of $\sigma$ is not fixed, since a fully Lorentz invariant description of the
noncommutative spacetime requires that {\sl all} values of~$\sigma$ which are
compatible with the spacetime uncertainty relations appear on the same footing,
i.e. as points in the set~$\Sigma$ of joint eigenvalues of the coordinates'
commutators. It is desirable to rid the Hamiltonian of the dependence
on~$\Sigma$. But since no  Lorentz  invariant average exists on~$\Sigma$, the
best one can do is the rotation invariant integration over~$\Sigma^{(1)}$, the
doubled sphere obtained when both the electric and magnetic part of~$\sigma$
have modulus one and hence are parallel, cf.~\cite{dfr}. 

The question of unitarity, however, is independent of whether the Hamiltonian
depends on $\Sigma$ or not, and in order to clarify the relation between the
Hamiltonian approach and the modified Feynman rules, we shall not perform
the integration over $\Sigma^{(1)}$ in this note\footnote{Note
however, that if the integration over $\Sigma^{(1)}$ is performed, the $\sigma$
variables in the various factors $H_I(t)$ in the perturbative expansion are
treated as {\sl independent} variables, whereas in the approaches outlined in
this note they are all identified with one another. Such points will be
discussed in detail elsewhere.}. In~\cite{BDFP} we will further investigate the
Hamiltonian approach.

From the Dyson series given in~\cite{dfr} we deduce that the corresponding
graph theory (for fixed~$\sigma$) again entails a planar and a nonplanar
contribution to the fish graph of $\phi^3$ theory, where the planar
contribution is again identical to the fish graph of ordinary quantum field
theory. The nonplanar contribution, however, differs from the one obtained in
the setting of the modified Feynman rules. This is due to the fact that in the
Hamiltonian approach the time ordering is performed with respect to the 
$t$~variables in each factor $H_I(t)$ of the perturbative expansion, such that the
resulting Heaviside functions are not involved in the nonlocal 
products~\cite[eqn. 6.15]{dfr}.  In fact, the full fishgraph in the Hamiltonian
approach is proportional to\footnote{We will comment on the proper definition
of Wick monomials in field theories on a noncommutative spacetime elsewhere.} 
\begin{eqnarray*} 
&&\hspace{-9pt}
\int dt_1dt_2
\hspace{-8pt}
\int\limits_{x_0=t_1} \hspace{-8pt} d^3x 
\hspace{-8pt}
\int\limits_{y_0=t_2} \hspace{-8pt}d^3y 
\,\,\theta(x_0-y_0)\cdot[\,
\langle p|:\! \phi(x)\phi(y)\!:|p\rangle
\stackrel{sym\;}{*_\sigma}\Delta_+(x-y)
\stackrel{sym\;}{*_\sigma}\Delta_+(x-y)\,]
\\
&&\hspace{-22pt}
+\,\int dt_1dt_2
\hspace{-8pt}
\int\limits_{x_0=t_1} \hspace{-8pt} d^3x 
\hspace{-8pt}
\int\limits_{ y_0=t_2} \hspace{-8pt}d^3y 
\,\,\theta(y_0-x_0)\cdot[\,
\langle p|:\! \phi(x)\phi(y)\!: |p\rangle
\stackrel{sym\;}{*_\sigma}\Delta_+(y-x)
\stackrel{sym\;}{*_\sigma}\Delta_+(y-x)\,]\,.
\end{eqnarray*} 
Here, $\stackrel{sym\;}{*_\sigma}$ stands for the symmetrized star product
with respect to both $x$ and $y$. Note that the Heaviside function $\theta$ is
multiplied {\sl pointwise} to the threefold star product and can in general
not be combined with the propagators $\Delta_\pm$ to yield the Feynman
propagator. For this reason and since 
\vspace{-10pt}$$
\overline{\langle p|\dots |p\rangle \stackrel{sym\;}{*_\sigma}
\Delta_\pm\stackrel{sym\;}{*_\sigma}\Delta_\pm}
\;=\;\langle p|\dots |p\rangle \stackrel{sym\;}{*_\sigma}\Delta_\mp
\stackrel{sym\;}{*_\sigma}\Delta_\mp\,,
$$
we can deduce that the unitarity condition at second order is indeed
satisfied, as 
\begin{eqnarray*} 
&&\theta\cdot[\,
\langle p|\dots |p\rangle
\stackrel{sym\;}{*_\sigma}\Delta_+
\stackrel{sym\;}{*_\sigma}\Delta_+\,]
\;+\;(1-\theta)\cdot[\,
\langle p|\dots|p\rangle
\stackrel{sym\;}{*_\sigma}\Delta_-
\stackrel{sym\;}{*_\sigma}\Delta_-\,]
\\
&+&\theta\cdot[\,
\overline{\langle p|\dots|p\rangle
\stackrel{sym\;}{*_\sigma}\Delta_+
\stackrel{sym\;}{*_\sigma}\Delta_+}\,]
\;+\;(1-\theta)\cdot[\,
\overline{\langle p|\dots |p\rangle
\stackrel{sym\;}{*_\sigma}\Delta_-
\stackrel{sym\;}{*_\sigma}\Delta_-}\,]
\\&=&
\langle p|\dots |p\rangle
\stackrel{sym\;}{*_\sigma}\Delta_-
\stackrel{sym\;}{*_\sigma}\Delta_-
\;+\;
\langle p|\dots |p\rangle
\stackrel{sym\;}{*_\sigma}\Delta_+
\stackrel{sym\;}{*_\sigma}\Delta_+\,,
\end{eqnarray*} 
the last expression being the one which also arises in the product of
two tree graphs.

The Hamiltonian approach, however, not only breaks Lorentz invariance
explicitly, but moreover does not seem to allow for  a simple definition of the
interacting field. A better way to define the latter, which actually has
already been investigated in the context of nonlocal theories
in~e.g.~\cite{kristensen,marnelius}, is to solve the field equation
perturbatively~\cite{yang}. As an illustrative example let us again consider a
massive scalar field with $\phi^3$ self-interaction.

Let $\phi=\sum g^n\,\phi_n$ be an expansion of the interacting field as a power
series with respect to the coupling constant. Then the field equation 
is 
$$
(\square-m^2)\, \phi_n = - \sum_{k=0}^{n-1}\phi_k*_{\sigma} \phi_{n-k-1}\,.
$$
Hence, $\phi_0$ is a free field. If it is identified with the incoming
field, then $\phi_1$ is given by 
$$
\phi_1=\Delta_{ret}\times (\phi_0*_{\sigma}\phi_0),
$$
$\times$ being the ordinary convolution, and for $\phi_2$ we obtain
\begin{eqnarray*}
\phi_2&=&\Delta_{ret}\times (\phi_0*_{\sigma}\phi_1+
\phi_1*_{\sigma}\phi_0)
\\&=&\Delta_{ret}\times (\phi_0*_{\sigma}(\Delta_{ret}\times 
(\phi_0*_{\sigma}\phi_0)))+
\Delta_{ret}\times((\Delta_{ret}\times 
(\phi_0*_{\sigma}\phi_0))*_{\sigma}\phi_0)\,.
\end{eqnarray*}

The once contracted terms in $\phi_2$ yield the fish graph, and again
we find two different contributions,
\begin{eqnarray}
\int dy \, \Delta_{ret}(y)\,(\Delta_+(y)+\Delta_-(y))\,\phi_0(x-y)
&&\quad\mbox{planar part}\label{YFfishpl}\\
\int dy \, \Delta_{ret}(y)\,(\Delta_+(y)*\phi_0(x-y) 
+\phi_0(x-y)*\Delta_-(y))
&&\quad\mbox{nonplanar part}\qquad\label{YFfishnpl}
\end{eqnarray}
Since the Moyal star product 
is not only strongly closed~\cite{co}, but also 
has the special property that one star product under an integral may be
replaced by a pointwise product, the nonplanar part is equal to 
\begin{equation}\label{YFfishnpl2}
\int dy \, (\Delta_{ret}* \Delta_+(y)+\Delta_-*\Delta_{ret}(y)) 
\,\phi_0(x-y) \,.
\end{equation}
The theory is unitary as long as the interacting field is
Hermitean, which is true by construction as long as $\phi_0$ is Hermitean.
In particular, Hermiticity is clearly fulfilled for the above expressions
(\ref{YFfishpl}) and~(\ref{YFfishnpl}),~(\ref{YFfishnpl2}). 
For the planar part this means 
\begin{eqnarray*}
\Delta_{ret}\,(\Delta_++\Delta_-)&=&(\theta\Delta)\,(\Delta_++\Delta_-)\;
=\;-\,i\,(\theta\Delta_+^2-\theta\Delta_-^2)\\
&=&-i\,(\theta\Delta_+^2+(1-\theta)\Delta_-^2)+i\Delta_-^2\\
&=&-\,i\,\Delta_F^2+i\Delta_-^2\\
&=&
+\,i\,\bar\Delta_F^2-i\Delta_+^2\,.
\end{eqnarray*}
Hence, for the planar part the Hermiticity condition at second order is 
identical to the unitarity condition~(\ref{uni}) for the
Feynman propagator in ordinary field theory.

The nonplanar part is Hermitean by construction as well. Obviously,
$\Delta_{ret}* \Delta_++ \Delta_-*\Delta_{ret}$ is its own complex
conjugate and in particular, 
\begin{eqnarray*}
\Delta_{ret}* \Delta_++ \Delta_-*\Delta_{ret}&=&
-\,i\,[\,\theta\Delta_+*\Delta_+-\theta\Delta_-*\Delta_+
+\Delta_-*\theta\Delta_+-\Delta_-*\theta\Delta_-\,]
\\&=& +\,i\,\Delta_-*\Delta_-
-\,i\,[\,\Delta_F*\Delta_++\Delta_-*\Delta_F-\Delta_-*\Delta_+\,]
\\&=&
+\,i\,\Delta_-*\Delta_--\,i\,\Delta_F*\Delta_F
-\,i\,[\,i\,\Delta_{ret}*\Delta_+-i\,
\Delta_{ret}*\Delta_F\,]
\\&=&+\,i\,\Delta_-*\Delta_--\,i\,\Delta_F*\Delta_F
+\,i\,\Delta_{ret}*\Delta_{av}
\\&=&-\,i\,\Delta_+*\Delta_++\,i\,\bar\Delta_F*\bar\Delta_F
-\,i\,\Delta_{av}*\Delta_{ret}\,.
\end{eqnarray*}
We conclude that the Yang Feldman approach modifies the ``scattering
amplitude'' of ordinary quantum field theory by the additional term
$\Delta_{ret}*\Delta_{av}$. It is precisely this term which renders the
theory unitary; as we have seen in~(\ref{nonunistar}), its absence in the
setting of the modified Feynman rules entails a violation of unitarity.

Again, if $\sigma$ is chosen such that there is a  time- or lightlike vector
$n$ with $\sigma^{\mu\nu}n_\nu=0$, we recover the unitarity condition
$\Delta_F*\Delta_F + \bar \Delta_F*\bar\Delta_F \;=\;
\Delta_+*\Delta_++\Delta_-*\Delta_- $.

We have thus seen that, as long as a proper perturbative setup is
employed, field theories on time/space noncommutative spacetimes may well be
unitary in the sense that probabilities are always conserved. The more subtle
problem of asymptotic completeness in such theories, which has also been
studied before in the context of nonlocal theories, will be pursued elsewhere.

\vspace{4ex}\noindent
{\bf Acknowledgement}

\noindent 
We would like to thank Bert~Schroer and Wolfhart~Zimmermann for valuable hints
to publications on nonlocal field theories.

\end{document}